\documentclass[%
 reprint,
superscriptaddress,
 amsmath,amssymb,
 aps,
]{revtex4-1}

\usepackage{color} 

\usepackage{ulem} 
\usepackage{graphicx}
\usepackage{caption}
\usepackage{subcaption}
\usepackage{dcolumn}
\usepackage{bm}
\usepackage{hyperref}

\begin{document}

\preprint{APS/123-QED}

\title{Solid Capillarity: When and How does Surface Tension Deform Soft Solids?}

\author{Bruno Andreotti}
\affiliation{Physique et M\'ecanique des Milieux H\'et\'erog\`enes, UMR
7636 ESPCI -CNRS, Universit\'e Paris-Diderot, 10 rue Vauquelin, 75005, Paris, France}

\author{Oliver B\"aumchen}
\affiliation{Max Planck Institute for Dynamics and Self-Organization (MPIDS), D-37077 G\"ottingen, Germany}

\author{Fran\c{c}ois Boulogne}
\affiliation{Department of Mechanical and Aerospace Engineering, Princeton University, Princeton, NJ 08544, USA}

\author{Karen E. Daniels}%
\affiliation{Department of Physics, North Carolina State University, Raleigh, NC 27695 USA}

\author{Eric R. Dufresne}%
\email[]{ericd@ethz.ch}
\affiliation{School of Engineering and Applied Sciences, Yale University, New Haven, CT 06520, USA}
\affiliation{Department of Materials, ETH Z\"urich, CH-8093 Zurich, Switzerland}

\author{Hugo Perrin}
\affiliation{Physique et M\'ecanique des Milieux H\'et\'erog\`enes, UMR
7636 ESPCI -CNRS, Univ. Paris-Diderot, 10 rue Vauquelin, 75005, Paris, France}

\author{Thomas Salez}
\affiliation{PCT Lab, UMR CNRS 7083 Gulliver, ESPCI ParisTech, PSL Research University, 75005 Paris, France}

\author{Jacco H. Snoeijer}%
\affiliation{%
Physics of Fluids Group, Faculty of Science and Technolog, and Burgers Center for Fluid Dynamics, University of Twente, 7500AE Enschede, The Netherlands
}
\affiliation{%
Mesoscopic Transport Phenomena, Eindhoven University of Technology, Den Dolech 2, 5612 AZ Eindhoven, The Netherlands
}

\author{Robert W. Style}
\affiliation{Mathematical Institute, University of Oxford, Oxford, OX1 3LB, UK}

\date{\today}

\begin{abstract}
Soft solids differ from stiff solids in an important way: their surface stresses can drive large deformations.
Based on a topical workshop held in the Lorentz Center in Leiden, this Opinion highlights some recent advances in the growing field of {\itshape solid capillarity} and poses key questions for  its advancement.
\end{abstract}

\pacs{Valid PACS appear here}
\maketitle

\begin{figure}
\centering
\includegraphics[width=5.7cm]{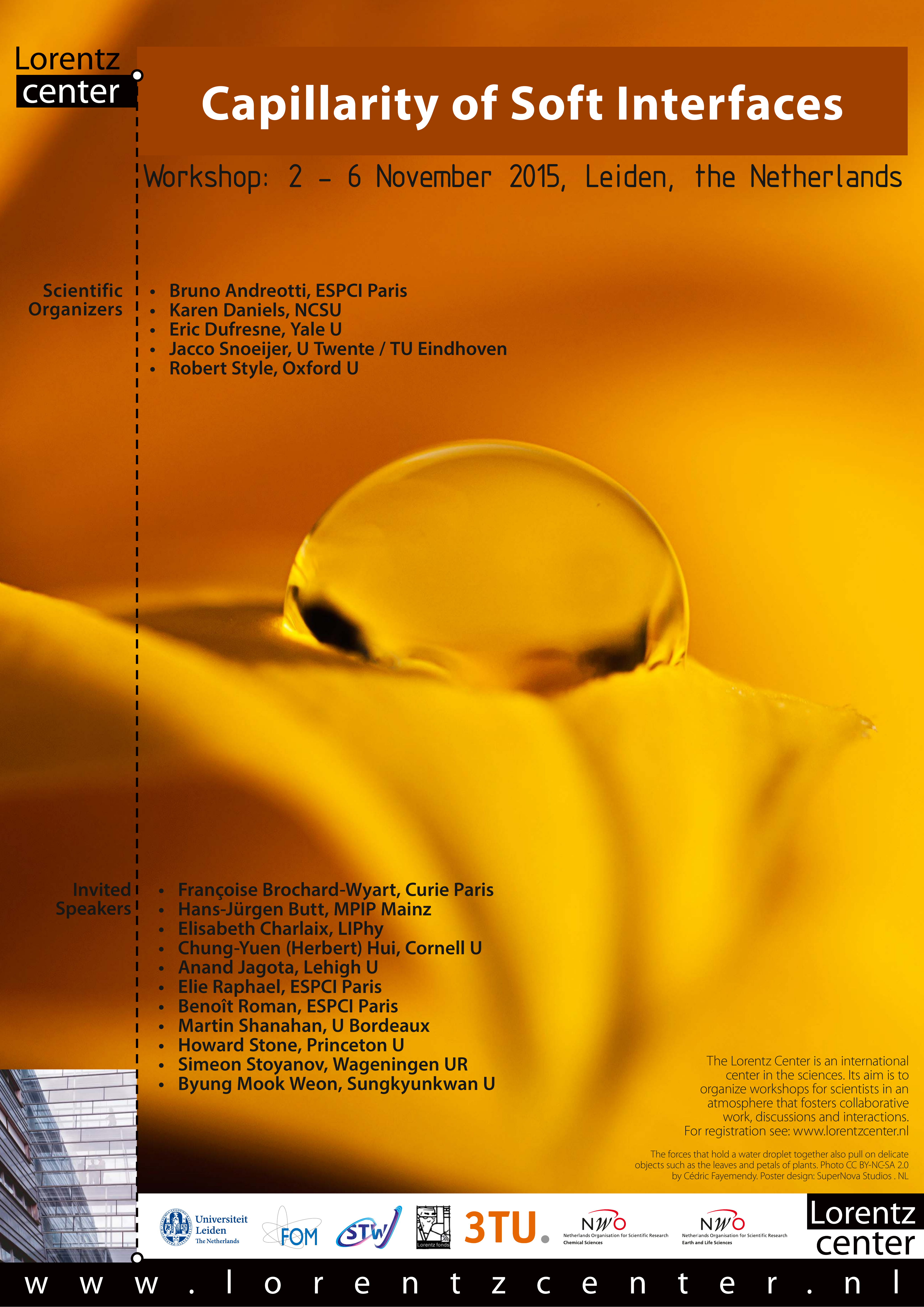}
\caption{A workshop on Capillarity of Soft Interfaces was hosted by the Lorentz Center at the University of Leiden in November 2015.  The poster for the event depicts a water droplet on a flower petal.}
  \label{fig:poster}
\end{figure}

Surface tension plays a dominant role in the mechanics of liquids at small length scales.
In Figure~\ref{fig:poster}, the water droplet curves into a spherical shape to minimize the energy due to the liquid-vapour interface. At the contact line, where the liquid, vapour, and solid meet, the orientation of the liquid interface minimizes the energy associated with all three interfaces, as described by the famous Young-Dupr\'{e} (YD) equation \cite{dege10}. Surface tension can also drive the motion of droplets, for example pulling them out of a tube \cite{Piroird2011} or propelling them along a surface \cite{Darhuber2005}.

Several recent studies have highlighted a new role for surface tension in the mechanics of deformable solids.
For instance, liquid droplets can bend slender solid objects \cite{shan87b,roma10} like the flower petal in Figure~\ref{fig:poster}.
Additionally, a solid's own surface tension can strongly couple to its deformation.
 For instance, soft cylinders can develop Rayleigh-Plateau pearling instabilities \cite{mora10} and soft composites can be stiffened by the surface tension of their inclusions \cite{styl15, Ducloue2014}.
The coupling of surface tension and deformation also plays an important role in wetting and adhesion, challengeing classic results like the Young-Dupr\'{e} equation and the JKR theory of contact mechanics \cite{john71}.

To foster the burgeoning field of solid capillarity and identify the most important new areas for investigation, the Lorentz Center at the University of Leiden hosted a workshop on
\href{https://www.lorentzcenter.nl/lc/web/2015/714/info.php3?wsid=714}{Capillarity of Soft Interfaces}
in November 2015.
In this Opinion, we highlight some of the key questions identified by the 49 workshop participants, including both academic and industrial researchers.
These issues center on the mechanical properties of soft solid interfaces and their implications for bulk deformation.


\section{Physics of Soft Solid Interfaces}

While the physical origins of the surface properties of stiff solids are well understood, a systematic investigation of the surface properties of soft solids remains to be undertaken.
The aims are not only to reveal the appropriate continuum-scale description of the mechanical properties of the interface, but also to understand their origins to allow for effective manipulation.

\subsection{\label{sec:shuttleworth} Continuum Description of Soft Interfaces}

{\it Key Question: What is the appropriate description of the mechanics of a soft solid interface?}

For a simple liquid, the energy gained from stretching the liquid-vapour interface is equal to the product of the surface energy $\gamma$ $[\rm J/\rm m ^2]$ and the gain of interfacial area.
Moreover, the surface stress, $\Upsilon$ $[\rm{N} /\rm{m}]$, is equal to the surface energy, and both are often referred to as  the surface tension.

For an elastic solid, the stress state of the interface is quite different.
As above, stretching the  interface increases the energy in proportion to the surface energy.
However, a second contribution arises when the surface energy is strain-dependent, leading to the so-called {\it Shuttleworth effect} \cite{shut50}.
In this case, the surface stress $\Upsilon$  not only depends upon the surface energy, $\gamma$, but also its derivatives with strain, as $\Upsilon=\gamma + {\rm d} \gamma / {\rm d} \epsilon$.
In contrast to liquids, the stress state of solids can be both anisotropic and strain-dependent.
For stiff solid materials,  some of the implications of this have been worked out \cite{ camm94b}.
For soft solids, the Shuttleworth effect is predicted to qualitatively change wetting phenomena, as discussed below.

An alternative perspective on the  state of stress at surfaces comes from the study of complex fluid interfaces.
When a fluid interface is laden with surfactants or particles, it can develop a  stress state that is anisotropic, strain-dependent,  and rate-dependent \cite{fuller2012}.
This surface rheology can have a significant impact on the bulk flow of fluids \cite{Langevin2000,hermans2015}.

To understand the surface mechanical properties of soft solids, it will be essential to integrate insights from  traditional solid mechanics (incorporating Shuttleworth's results) with emerging insights from the surface rheology of complex fluid interfaces. Whatever the results,  new experimental approaches are needed to quantify the surface stresses of soft solids.

\subsection{\label{sec:surface} Physicochemical Origins of Surface Properties}

{\it Key Questions:  What is the origin of the surface mechanical properties? How can they be manipulated?}

Polymeric materials have been the dominant choice for experiments in the mechanics of soft solids.
These materials feature a system-spanning cross-linked polymer network; this network may be  swollen either by un-crosslinked polymer, or by a chemically-distinct solvent.
This  general description encompasses a  broad class of soft materials, including polymeric elastomers and gels, with a wide range of properties arising from the particular chemistry.
In all cases, the surface properties will be impacted by the structure of the polymer network near the interface.
For gels, the solvent and additional surface-active components may significantly impact the surface properties.

We can learn a lot about possible contributions of the polymer network from our emerging understanding of a closely-related class of materials: polymer melts.
Polymer melts are composed of long entangled polymer chains, without cross-linkers, and are known to have  significantly-modified  mechanical properties near their surfaces.
The presence of interfaces and confinement may alter the network of entanglements of thin polymer films, thus reducing their resistance to deformation \cite{bodi06,shin07,How08,Baumchen2009}.

Surfaces can also modify the thermodynamic properties of polymeric materials.
The glass-transition temperature of a polymer melt can be significantly reduced in confinement, in a manner that  sensitively depends on the boundary conditions  \cite{kedd94,forr96,Baumchen2012}.
Furthermore, experiments recording the evolution of surface perturbations have shown that the surface mobility of a glassy polymer can be significantly enhanced relative to the bulk \cite{Fakhraai2008,yang10,Chai2014}.

Are these phenomena manifest in softer  materials such as gels?
How do the structure and properties of these materials vary near the surface?
Moving forward, we need to quantify changes in the structure and composition of soft solids in the vicinity of interfaces, and correlate these findings with measurements of surface mechanical properties. With these insights, we will be able to rationally design the mechanical properties of soft interfaces.

\section{\label{sec:applications} Coupling Surface Stresses to Bulk Deformation}

Surface stresses can dramatically impact the way soft solids deform \cite{jago12}.
These solid capillary effects become significant below a critical {\itshape elastocapillary length}  $L = \Upsilon/E$, where $E$ is the elastic modulus.
The basic physics is highlighted by the following argument:
Consider a surface with a sinusoidal corrugation of wavelength, $\lambda$, and amplitude, $A$.
Surface stresses act to flatten the surface, with  a  stress given by the Laplace law that scales like $ \Upsilon A/\lambda^2$.
At the same time, elastic forces will resist this deformation with a restoring stress that scales like $E A/\lambda$.
When $\lambda \ll L$, surface stresses overpower elastic restoring forces, thereby flattening the surface.
The elastocapillary length is on the micron-scale for gels, nanometre-scale for elastomers, and is irrelevant for structural materials.
Here, we discuss two applications of the competition between surface and bulk effects in soft solids: wetting and adhesion.

\subsection{Wetting}

{\it Key Questions: How do strain-dependent surface energies impact wetting and other elastocapillary phenomena? What controls dynamic elastocapillary phenomena?}

Classically, wetting is a pure surface effect, governed by the surface energies of the three phases \cite{dege10}.
Recently, it has been shown that deformation of soft substrates by liquid surface tension \cite{Pericet-Camara2008,Jerison2011} can have a dramatic impact on the essential phenomena of wetting, including departures from the Young-Dupr\'e law \cite{Lubbers2014, styl12,Bostwick2014,Limat2012}.

The majority of the experimental work on soft wetting has presumed that the surface stress of a soft solid is liquid-like (isotropic and strain independent).
As discussed above, this is  not a robust assumption for soft solids or complex fluids.

Recent theoretical work has shown that the  state of stress near a three-phase contact line should be significantly changed when a surface is subject to the Shuttleworth effect, {\itshape i.e.} the substrate has a strain-dependent surface energy, $\gamma(\epsilon)$ \cite{Weijs2013,Neuk2014}. The equilibrium state  minimizes  the sum of the surface and strain-induced elastic energies.
Here, the strain-dependence of the surface energy induces an additional elasto-capillary coupling, which for a drop on a membrane can lead to very large strains near the contact line  \cite{Hui2015b,schu15}.
For a thin rod that is partially immersed, the Shuttleworth effect leads to a discontinuity in strain \cite{Weijs2014}, suggesting that great care must be taken when applying the classical capillary ``force'' picture near the contact line. These ideas are consistent with an experiment reported in \cite{marc12}, but further experimental investigation is needed.

While some progress has been made on the importance of simple departures from ideal fluid-like surface tensions in the case of wetting, the issue is completely unexplored in other elastocapillary phenomena. For instance, what are the implications of complex surface rheology in adhesion, fracture mechanics, and composites?

So far, the study of elastocapillary phenomena has focused on static situations.
However, the dynamics of these phenomena are ripe for investigation.
In the context of wetting, the deformation of the substrate due to liquid surface tension creates a new source of dissipation.
This dissipation, called {\it viscoelastic braking}, can dominate the motion of a contact line, causing droplets to slide more slowly on a soft surface than a stiff one \cite{shan95,long96}.
Experiments and modelling have directly related the speed of an advancing contact line to substrate viscoelasticity \cite{kaji2013,karp15}.

On the other hand, coupling of surface stresses to bulk deformation can also {\it drive} contact-line motion.
For example, droplets spontaneously translate down gradients in the stiffness of soft substrates \cite{styl13b}.
In this case, however, we do not have a physical model to predict the speed of droplet motion.

More generally, the dynamics of contact lines on soft surfaces could be modified by a variety of dissipative phenomena including poroelastic flow and substrate plasticity.
A broad range of exploratory experiments are needed.

\subsection{\label{sec:adhesion} Adhesion}

{\it Key Question: How does solid-vapour surface tension modify the adhesive contact of elastic solids?}

The standard model of adhesive solid contacts was developed by Johnson, Kendall, and Roberts (JKR) \cite{john71}.
It features a competition of surface adhesive and bulk elastic energies, captured by the dimensionless parameter, $ER/W$.
Here, $R$ is a length scale associated with a contact and $W$ is the work of adhesion.
However, it does not include contributions from the surface stress of the solid-vapour interface, captured by the elastocapillary parameter, $ER/\Upsilon$.

In the limit of small contacts on soft substrates, theoretical analyses suggest that the JKR theory will be replaced by the Young-Dupr\'{e} condition with the soft substrate in the place of the wetting liquid \cite{carr10,styl13c,sale13,xu14}.
Recently, a JKR-YD crossover was observed using silicone gels~\cite{styl13c,Jensen2015} at a length scale around $10~\mu$m.
However, the generality of these effects needs to be established with further experiments.
First, the JKR-YD crossover should be explored for a wider range of materials including hydrogels and recently-developed monophasic ultrasoft materials \cite{dani2015}.
 The effect of solid-vapour surface tension should also impact nanoscale contacts on stiffer polymeric materials, like elastomers \cite{Lau2002}.
At this scale, long-range forces, embodied by the Derjaguin-Muller-Toporov theory \cite{derj87}, may also compete with surface tension.

The role of solid-vapour surface tension in solid-solid contact mechanics could become manifest at larger length scales  by adjusting another parameter: the thickness, $h$, of the elastic medium.
If $h \ll R$, a separation of length scales leads to the dominance of much softer elastic modes, associated with the stretching modulus
or bending rigidity.
This geometric effect is already being exploited in studies of wetting, where liquid surface tension creates macroscopic deformations of partially-wetted rods and sheets ~\cite{roma10}.
We anticipate that adhesion between slender solid objects will be a productive avenue for further investigation.

Studies of the impact of solid-vapour surface tension on adhesive contacts have focused on the special case of zero load in equilibrium.
However, it is essential to understand the effects of external loading and the dynamical response.
For instance, the pull-off force and peeling dynamics are expected to depart dramatically from the pure elastic case~\cite{Hui2015}. In the same spirit, following  recent developments \cite{liu14}, the singularity of fracture mechanics near a soft crack tip should be regularized by surface tension. Once again, there is a direct call for experiments to test those new fundamental ideas -- with numerous practical implications.

\section{Conclusions}

The study of solid capillarity is in its early stages.
Fundamental challenges remain in measuring and manipulating the surface mechanical properties of soft solids.
The coupling of surface stresses to bulk deformation presents a diverse range of problems, in wetting, adhesion, and other interfacial phenomena involving soft solids.
This nascent field will benefit from cross-fertilization with solid mechanics, rheology, and polymer science.
Coupling between surface stress and bulk deformation also plays an important role in many biological phenomena \cite{mann10,mert12,gonz12,murr14}, and robust dialog between the physical and biological communities promises to be very fruitful.

\section{Acknowledgments}

We thank all the participants at the workshop and  the staff of the Lorentz Center for a productive and thought-provoking week. The workshop was sponsored by the 3TU Center of Competence FSM (Fluid and Solid Mechanics) and ERC (the European Research Council) Consolidator Grant No. 616918.
F.B. acknowledges financial support from the People Programme (Marie Curie Actions) of the European Union's Seventh Framework Programme (FP7/2007-2013) under REA grant agreement 623541.
O.B. acknowledges support from the ESPCI Joliot Chair

%

\end{document}